\documentclass[12pt]{article} 
\usepackage{algorithm}
\usepackage{algorithmic}

\usepackage{amsmath}
\usepackage{graphicx}
\usepackage{enumerate}
\usepackage{natbib}
\usepackage{multirow}
\usepackage{longtable}
\usepackage{booktabs}
\usepackage{amssymb}
\usepackage{bm}
\usepackage{mathrsfs}
\usepackage{amsthm}
\usepackage{amsfonts}
\usepackage{subfigure}
\usepackage{epsfig,amssymb,latexsym,verbatim}
\usepackage{epstopdf}
\usepackage{graphics}
\usepackage[english]{babel}
\usepackage[figuresright]{rotating}
\usepackage[dvipsnames]{xcolor}
\usepackage{color}
\usepackage{hyperref}

\usepackage{xr}

\allowdisplaybreaks
\newtheorem{theorem}{Theorem}

\newtheorem{remark}{Remark}

\numberwithin{equation}{section}
\numberwithin{lemma}{section}
\numberwithin{theorem}{section}
\numberwithin{prop}{section}
\numberwithin{corollary}{section}

\DeclareMathOperator*{\argmax}{arg\,max}
\def\P{\mathbb{P}}
\def\E{\mathbb{E}}
\def\Var{\mbox{Var}}
\def\Cov{\mbox{Cov}}
\def\o{{\scriptstyle{\mathcal{O}}}}
\def\O{\mathcal{O}}

\newcommand{\blind}{0}

\begin{document}

\def\spacingset#1{\renewcommand{\baselinestretch}%
{#1}\small\normalsize} \spacingset{1}

\date{}
\if0\blind
{

           
  \title{\bf Recursive Multiple Change Point Detection of Nonstationary Time Series: Instability Tests, Estimation and Confidence Intervals}
	\author{Leheng Cai$^1$,   and Zhou Zhou$^2$
		\\~\\
		{\small \it $^{1}$  Department of Statistics and Data Science, Tsinghua University}
		\\{\small \it $^{2}$ Department of Statistical Sciences, University of Toronto}
	}
} 

\if1\blind
{
  \bigskip
  \bigskip
  \bigskip
  \begin{center}
    { \bf }
\end{center}
  \medskip
} \fi

\bigskip

\maketitle
	
	\textbf{Abstract:}   
We develop {\textbf b}ootstrap-{\textbf a}ssisted {\textbf r}obust {\textbf b}inary {\textbf s}egmentation (BARBS), a recursive binary segmentation method for multiple change point   detection under general nonstationary temporal dynamics. A  novel  Gaussian multiplier bootstrap for the CUSUM statistics is proposed, offering robustness to complex dependence structures.
Through meticulous calibration of the critical values at each stage of the recursion, BARBS ensures   control of the  Type I error under the null hypothesis of no change points. 
When change points are present, BARBS identifies the correct number of changes with a prespecified  probability, and the resulting change point location estimators attain the same uniform consistency rate as classical binary segmentation. 
 Building on this,  we introduce second-stage refined estimators that achieve the optimal individual localization rate,  and  establish their asymptotic distributions and nearly
optimal uniform localization rates  under both fixed and vanishing jump magnitudes. Extensive numerical experiments across various settings confirm the robustness and superior performance of BARBS relative to existing approaches. To illustrate the practical relevance of the proposed methodology, we analyze U.S. inflation data, yielding change points that align with several documented macroeconomic episodes. 
 

	\maketitle
	
\newpage
\spacingset{1.5} 

\section{Introduction}

Change point detection is a fundamental problem in modern data science, where the goal is to identify points in time at which certain distributional properties of the underlying data generating process undergo abrupt changes. 
For standard treatments of change point analysis, we refer to  \cite{csorgo1997limit} and \cite{horvath2024change}. 
Such structural shifts  or change points  may reflect significant events or transitions in the observed system, such as economic regime changes \citep{hamilton1989new, bai1998estimating}, shifts in climate patterns \citep{elsner2008increasing, jandhyala2010change}, or applications in medical care and biological research   \citep{clifford2015physionet, wu2024frequency}. Accurately detecting and localizing these changes is crucial for effective modeling, forecasting, and decision-making in a wide range of scientific and industrial applications.

Consider the observed time series $\{X_{i}\}_{i=1}^n$, which follows the model
\begin{align}
	X_{i} = \beta(i/n) + \varepsilon_i, \quad i = 1, \dots, n,
\end{align}
where the trend $\beta(\cdot)$ is a bounded and piecewise constant function with $r_n\geq 0$ change points. Specifically, there exist breakpoints $0 = \tau_0 < \tau_1 < \cdots < \tau_{r_n} < \tau_{r_n+1} = 1$ such that $\beta(i/n) = \beta_l$ for $\tau_l < i/n \leq \tau_{l+1}$. Denote by $\Delta_l=\beta(\tau_l+)-\beta(\tau_l-)$ the jump magnitude. The number of change points $r_n$ is allowed to diverge as the sample size $n \to \infty$.  
The error process $\{\varepsilon_i\}_{i=1}^{n}$ is a nonstationary sequence with mean zero, whose covariance and higher-order dependence structures may vary over time. A more detailed framework describing the temporal dependence is provided in  Section~\ref{sec:methodology}. 
Nonstationarity is often the rule rather than the exception in modern
temporally ordered data \citep{dahlhaus1997fitting}. In many scientific, engineering, economic, environmental, and biomedical applications, the variance, serial dependence, and other characteristics of the underlying dynamics evolve over time, either gradually or through abrupt changes \citep{karmakar2022simultaneous,sansom2019state,sanderson2010estimating}. The temporal evolution in the error process and
the structural changes in the mean that are of primary scientific interest can be easily confounded by procedures calibrated under
stationarity \citep{zhou2013heteroscedasticity,gorecki2018change}, thus ignoring such nonstationarity may result in misleading scientific conclusions.

A fundamental and important task is to  test for the presence of change points in the trend over time, and further accurately estimate the number $r_n$ of change points and   their locations $\tau_1,\ldots, \tau_{r_n}$    if they exist. 
The classical literature on multiple change point detection can be broadly categorized into three main methodological frameworks: loss-based methods, moving sum (MOSUM)-based methods, and binary segmentation-based methods.
Loss-based methods typically estimate the underlying piecewise constant signal by minimizing a loss function, often derived from the likelihood or quasi-likelihood, together with a regularization penalty. For example, \cite{killick2012optimal} proposed the pruned exact linear time (PELT) algorithm, which achieves linear computational complexity in the sample size n under independent and identically distributed (i.i.d.) noise. Extending such methods to nonstationary time series with complex dependence, however, remains challenging. Moving sum (MOSUM)-based methods were first introduced by \cite{antoch2000change} and subsequently established on a rigorous theoretical foundation by \cite{eichinger2018mosum}. Despite their appealing computational efficiency, practical implementation requires the selection of a bandwidth parameter, and both bandwidth choice and the theoretical analysis become considerably more delicate in the presence of genuinely nonstationary noises.
 
Another typically used method for multiple change point detection is binary segmentation \citep{vostrikova1981detecting}.
Binary segmentation is widely appreciated for its computational efficiency, which splits the data into two segments based on an initial change point estimate, and then recursively applies the same procedure to each resulting segment until no further change points are detected.  When the number of change points is fixed, \cite{venkatraman1992consistency} established the consistency of binary segmentation under i.i.d. Gaussian noises; \cite{bai1997estimating} extended the method to linear models and   provided the convergence rate and asymptotic distribution of the binary segmentation change point estimators, which assumes the error  process is a linear process of a martingale difference sequence with constant variance.  When the number of change points diverges, \cite{Fryzlewicz_2014} proposed the wild binary segmentation (WBS) procedure under i.i.d. Gaussian errors, which is capable of detecting change points separated by short intervals and exhibiting small jump magnitudes. Building on this idea, \cite{baranowski2019narrowest} introduced the narrowest-over-threshold (NOT) method. More recently, \cite{kovacs2023seeded} proposed seeded binary segmentation (SBS), which substantially improves computational efficiency while enhancing reproducibility. Despite these advances, most existing methods are developed under the assumptions of i.i.d. Gaussian errors or stationary linear processes, thereby limiting their applicability to nonstationary time series and other settings with more complex data-generating mechanisms.

A key challenge in binary segmentation is deciding when to terminate the procedure, typically addressed through the following two approaches: hypothesis testing and thresholding aided by model selection. Among works focusing on hypothesis testing, \cite{bai1997estimating} proposed an F-test under the assumption of a stationary error process. However, this approach does not handle the multiple testing issue inherent in sequential procedures. 
On the other hand, the aforementioned methods such as \cite{Fryzlewicz_2014,baranowski2019narrowest}  employ stopping criteria based on thresholding, which typically rely on mathematically abstract  orders. In practice, such thresholds are often difficult to calibrate, and model selection procedures based on the i.i.d. assumption, often involving modifications of existing information criteria such as the Schwarz criterion, are typically required. However, the theoretical properties of these model selection procedures under complex dependence structures remain largely unexplored.

To the best of our knowledge, the literature on multiple change point detection under complex temporal dynamics remains sparse. Existing methods are largely built upon MOSUM-type procedures \citep{mies2023functional,wu2024multiscale,kohne2025edge,bai2026complex}. Furthermore, the nonstationarity considered in these studies is inherently structured: the error process is typically assumed to be piecewise locally stationary or generalized locally stationary under appropriate regularity conditions, implying that the underlying data-generating mechanism evolves smoothly within a finite or slowly diverging number of segments. Developing multiple change point detection procedures that transcend the MOSUM paradigm and accommodate genuinely nonstationary noise, in which the data-generating mechanism may vary arbitrarily over time subject only to weak dependence and moment assumptions, therefore remains an important and largely open problem.

In this paper, we fill this gap by  proposing {\bf bootstrap-assisted robust binary segmentation (BARBS)}, a novel recursive procedure for multiple change point detection under a genuinely nonstationary environment. The proposed method employs a multiplier bootstrap to calibrate the critical values of the CUSUM test statistics under complex temporal dependence, thereby avoiding reliance on pivotal Brownian bridge asymptotics. A related bootstrap idea was considered by \cite{yu2021finite} for testing change points in high-dimensional mean vectors under independent observations, though without adjusting for the significance level. In contrast, our procedure is designed for general nonstationary and dependent time series and incorporates a significance-level adjustment that enables reliable recursive change point detection. BARBS stands in contrast to the rough thresholding methods used in \cite{Fryzlewicz_2014} and \cite{baranowski2019narrowest}, which require additional procedures for model selection. 
Besides, we meticulously adjust the significance levels and critical values used in the sequential bootstrap tests at different stages of BARBS to   control  the Type I error when no change points are present, while accurately identifying the number of change points with the prespecified probability when they exist. The uniform convergence rate of change point estimators is also established. 
Additionally, building on the initial estimates obtained from BARBS, we propose  second-stage refined estimators that achieve the optimal individual localization rate. The asymptotic distributions and  nearly optimal uniform convergence rates of these refined estimators are derived  under both fixed jump magnitudes and those that shrink to zero, providing theoretical guarantees  for constructing confidence intervals for the change point locations.   Results on sequential Gaussian approximation under nonstationary time series in Lemma A.1 of the Supplementary
Materials, the recently developed anti-concentration inequality and Gaussian comparison result from \cite{wu2024frequency}, and properties of the physical dependence measure from \cite{wu2005nonlinear,liu2013probability,zhou2014} are interwoven seamlessly to achieve the satisfactory theoretical results.

\par This paper is organized as follows: Section \ref{sec:methodology} introduces the BARBS, and Section \ref{sec:theory} establishes    corresponding asymptotic properties. Section \ref{sec:repartition} introduces the second-stage refined change point estimators and further investigates   corresponding asymptotic behavior.    Numerical simulations  and a data application  are presented in Sections \ref{sec:simulation} and \ref{sec:realdata}, respectively. Additional simulation results and technical proofs of theoretical results are included in the Supplementary Materials. 

\section{Methodology}\label{sec:methodology}

\subsection{Preliminary}
Throughout the paper, let \(k_i^\ast=\lfloor n\tau_i\rfloor\) denote the corresponding integer-valued change point locations, and 
the $\mathcal L_q$ norm of a   random variable $X$ is
defined as $\left\|X\right\|_{q}=\left(\E|X|^q\right)^{1/q}$.  For two sequences $\{a_n\}$ and $\{b_n\}$, we write   $a_n\gtrsim b_n$ and $a_n\asymp b_n$ to mean 
that there exist   $c,C>0$  such that $a_n\geq  cb_n$ for all $n$, and $cb_n\leq a_n\leq Cb_n$ 
respectively.

To quantify the complex temporal dependence structure of the  nonstationary process $\{\varepsilon_i\}_{i=1}^{n}$, we employ the framework of physical dependence measure \citep{wu2005nonlinear}, and define $\varepsilon_i=\mathcal G_i(\mathcal F_i)$, where $\mathcal G_i$ is a measurable function, $\mathcal F_i=(\cdots,e_{i-1},e_i)$  is a  filtration, and $\{e_i\}_{i\in\mathbb Z}$ are i.i.d. random variables that generate the filtration.  Notably, the physical mechanism process $\mathcal G_i$ is allowed to vary with different $i$, which accommodates a broad class of nonstationary processes. The  
 dependence measure  is then defined as   $\delta_{k,q}=\max_{1\leq i\leq n}\left\|\mathcal G_i(\mathcal F_i)-\mathcal G_i(\mathcal F_{i,i-k})\right\|_q$, where     $\mathcal F_{i,i-k}=(\cdots,e_{i-k}^\prime,\cdots,e_i)$, and $e_{i-k}^\prime$ is an i.i.d. copy of $e_{i-k}$. Intuitively,  it encapsulates the uniform effect over time of $k$-lagged innovation in the underlying data generating process (DGP) on the current observation. 
\begin{remark}\label{remark1}
    One special example of such nonstationary time series is the piecewise locally stationary (PLS) model with $d$ break points. The process $\{\varepsilon_i\}_{i=1}^{n}$ is PLS($d$) if there exist constants $0=s_0<s_1<\cdots<s_d<s_{d+1}=1$ and $d+1$ measurable functions $\mathcal G_0,\ldots,\mathcal G_d$ such that \begin{align*}
        \varepsilon_i=\mathcal G_j(i/n,\mathcal F_i),\,\,\, s_j<i/n\leq s_{j+1}, \,\,\,j=0,1,\ldots,d.
    \end{align*}
    The time series is assumed to be locally stationary between change points $s_j$ and $s_{j+1}$ in the sense that the data generating mechanism evolves gradually within each segment, satisfying that 
        $\max_{0\leq j\leq d}\left\|\mathcal G_j(s,\mathcal F_0)-\mathcal G_j(t,\mathcal F_0)\right\|_q\leq C_{\mathcal G}|s-t|$ for some constant $C_{\mathcal G}>0$.
However, abrupt changes in the underlying mechanism are permitted at locations $s_j$, $j = 1, 2, \dots, d$, reflecting sudden shifts in the filtering process from $\mathcal G_{j-1}$ to $\mathcal G_j$. This PLS  framework captures both the smooth variation within segments and the abrupt structural breaks across segments, making it well-suited for modeling nonstationary noise  observed in real world data. 
A specific example  of PLS is illustrated in Section \ref{sec:simulation}, and further detailed discussions   of PLS processes and their associated dependence measures can be found in  \cite{zhou2013heteroscedasticity} and \cite{wu2024multiscale}.
\end{remark}

\subsection{Bootstrap-assisted testing and estimation}\label{sec:bootstrap}
To formalize our binary segmentation method more rigorously, suppose that at a certain stage of the procedure, we are working with a subsample that starts at index $s$ and ends at index $e$ such that $1\leq s<e\leq n$. Denote by $R=R_{s,e}$ the number of change points between $s$ and $e$.  
Consider the hypothesis testing problem that \begin{align}\label{H0}
    H_0:R=0\,\,\,{\text{vs.}}\,\,\, H_1:R\geq 1.
\end{align}
The following  CUSUM  test statistic is used for testing change point on the interval $[s,e]$.  
\begin{align}
	T_{s:e,m}&=\max_{s+m\leq k \leq e-m }\left|S_{s,k,e}\right|\label{teststat1},\\
	S_{s,k,e}&=\frac{1}{\sqrt{e-s+1}}\left(\sum_{i=s}^{k}X_i-\frac{k-s+1}{e-s+1}\sum_{i=s}^{e}X_i\right),\label{teststat2}
\end{align}
where $m$ is a tuning parameter satisfying $m\to\infty$ and $m/(e-s+1)\to 0$.

Note that (\ref{teststat2}) is the classic CUSUM test statistic used in \cite{zhou2013heteroscedasticity} for the subsample $\{X_i\}_{i=s}^{e}$. 
Under the null in (\ref{H0}), when $\{\varepsilon_i\}_{i=1}^{n}$ is a generally nonstationary time series, the limiting distribution of $T_{s:e,m}$ is very complex. In order to estimate the critical values for this test, we propose the following  multiplier bootstrap procedure to approximate the distribution of $T_{s:e,m}$. For any $s+m\leq k\leq e-m$, define 
\begin{align*}
      \mathcal S_{s,k,e}^{(b)} = \frac{1}{\sqrt{e-s-2m+1}}\left(\sum_{j=s+m}^{k}Y_jG_j^{(b)}-\frac{k-s-m+1}{e-s-2m+1}\sum_{j=s+m}^{e-m}Y_jG_j^{(b)}  \right), 
    \end{align*}
where $G_j^{(b)}$ are i.i.d.\ standard Gaussian random variables for the $b$-th bootstrap iteration ($1\leq b\leq B$), and \begin{align}\label{def:Yj}
    Y_j&= \frac{1}{\sqrt{2m}} \left(\sum_{i=j-m+1}^j X_i-\sum_{i=j+1}^{j+m}X_i \right),\;\;\; m\leq j\leq n-m.
\end{align}
The final bootstrap statistic  is defined by 
\begin{align*}
	\mathcal T_{s:e,m}^{(b)}=\max_{s+m\leq k\leq e-m }\left|\mathcal S_{s,k,e}^{(b)}\right|,\,\,\, 1\leq b\leq B.
\end{align*}
For some $\widetilde\alpha\in(0,1)$ depending on $[s,e]$, denote by $\text{crit}_{1-\widetilde\alpha}(\mathcal T_{s:e,m})$ the $(1-\widetilde\alpha)$-conditional quantile of $\mathcal T_{s:e,m}^{(b)}$ given the data $\{X_i\}_{i=1}^{n}$. 
One rejects the null hypothesis of no change point in $[s,e]$ at the significance level of $\widetilde\alpha$ if \begin{align}\label{rule}
    T_{s:e,m}>\text{crit}_{1-\widetilde\alpha}(\mathcal T_{s:e,m}). 
\end{align}   
In order to empirically generate a critical value for $T_{s:e,m}$, one takes 
\begin{align}\label{def:crit}
	\widehat{\text{crit}}_{1-\widetilde\alpha}(\mathcal T_{s:e,m})&=\text{Quantile}\left(\{\mathcal T_{s:e,m}^{(b)}\}_{b=1}^B,1-\widetilde\alpha \right).
\end{align}

If one rejects the null hypothesis  in (\ref{H0}),   the following weighted CUSUM-based statistic is adopted for  change point location:  
\begin{align}\label{def:hatk}
    \widehat k_{s:e} = \argmax_{s+m\leq k\leq e-m} \sqrt{\frac{(e-s+1)^2}{(k-s+1)(e-k)}}|S_{s,k,e}|.
\end{align}
\begin{remark}
  Here, if one directly uses the argmax of the absolute value of the standard CUSUM test statistic $S_{s,k,e}$ in (\ref{teststat2}) to estimate the change point location, the resulting estimator may not be consistent.  In certain multiple change point settings, the expectation of the standard CUSUM statistic may have a flat peak, preventing its maximizer from concentrating around a true change point; see Example 2.3.1 in \cite{horvath2024change} for instance.
\end{remark}

\subsection{The BARBS Procedure}\label{sec:bsprocedure}

\begin{figure}[h!]
    \centering
    \includegraphics[width=0.75\linewidth]{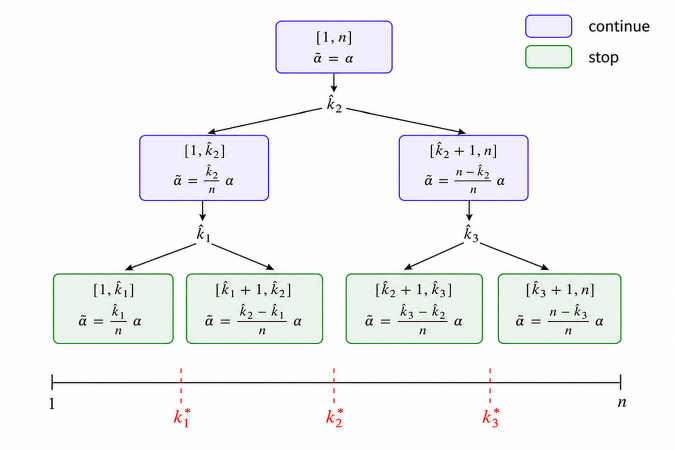}
    \caption{Illustration of the recursive testing in BARBS.}
    \label{fig:alpha-allocation}
\end{figure}

In BARBS, the overall significance level $\alpha$ is treated as a global error budget and is allocated across the recursive tests in proportion to interval length. Specifically, when the algorithm examines an interval $[s,e]$, the test in \eqref{rule} and \eqref{def:crit} is conducted at the local significance level $\widetilde{\alpha}
={(e-s+1)}\alpha/{n}$.
Figure~\ref{fig:alpha-allocation} illustrates how the significance budget is inherited through recursive splitting.
Intuitively, each observation is assigned an equal share $\alpha/n$ of the global error budget, so that an interval receives a budget proportional to the number of observations it contains.

The key observation is that, along the ideal recursive path, a false positive can occur only when BARBS rejects on an interval that contains no true change point and on which the recursion should therefore terminate. These terminal intervals lie between consecutive true change points, are mutually disjoint, and have total length at most $n$.  
Consequently, the sum of their local significance levels is bounded by $\alpha$.
 In this sense, the allocation in BARBS can be viewed as a length-weighted Bonferroni correction, which is crucial for asymptotically controlling the overall false positive probability at the nominal level $\alpha$. 
 
 Moreover, under the short-range dependence assumption imposed on the time series, the CUSUM tests conducted on these mutually disjoint terminal intervals are asymptotically independent. This asymptotic independence implies that the conservativeness introduced by Bonferroni is only of second order in the target significance level.

Algorithm~\ref{algorithm1} summarizes the BARBS procedure and returns a set \(\mathcal C\) of estimated change points. Accordingly, the estimated number of change points is \(\widehat r_n=|\mathcal C|\). We denote the ordered elements of \(\mathcal C\) by
$
\widehat k_1<\cdots<\widehat k_{\widehat r_n}$,
which are the estimated change point locations.

\begin{algorithm} 
\caption{Bootstrap-assisted robust binary segmentation (BARBS)}
\begin{algorithmic}[1]
\REQUIRE Time series data $\{X_i\}_{i=1}^n$,   window size $m$, significance level $\alpha$, and  minimum interval length $L$ with default value $2m+1$. 

\STATE Initialize $ {\mathcal{C}} \leftarrow \emptyset$
\STATE Initialize segment list $\mathcal{S} \leftarrow \{(1,n)\}$
\WHILE{$\mathcal{S}$ is not empty}
    \STATE Select and remove a segment $(s,e)$ from $\mathcal{S}$
    
        \STATE Compute test statistic $T_{s:e,m}$  in  (\ref{teststat1})
        \STATE Compute $\widehat{k}_{s:e}$  in  (\ref{def:hatk})
        \STATE Compute $\widehat{\text{crit}}_{1-\widetilde\alpha}(\mathcal T_{s:e,m})$ according to (\ref{def:crit}) with $\widetilde\alpha =  {(e - s + 1)\alpha}/{n}$  
        \IF{$T_{s:e,m} > \widehat{\text{crit}}_{1-\widetilde\alpha}(\mathcal T_{s:e,m})$}
        \STATE Add change point $\widehat{k}_{s:e}$ to $ {\mathcal{C}}$
        \IF{$\widehat{k}_{s:e} - s + 1 \geq  L$}
           \STATE   Add  new segment  $(s,\widehat{k}_{s:e})$  to $\mathcal{S}$
        \ENDIF
        \IF{$e-\widehat{k}_{s:e}    \geq  L$}
            \STATE Add  new segment   $(\widehat{k}_{s:e}+1,e)$ to $\mathcal{S}$
         \ENDIF
    \ENDIF
\ENDWHILE
\ENSURE A set of estimated change points $ {\mathcal{C}}$
\end{algorithmic}
\label{algorithm1}
\end{algorithm}

\section{Theoretical results} \label{sec:theory}

To investigate asymptotic properties of the BARBS procedure, the following mild assumptions are introduced.  
\begin{itemize}
    \item[(A1)] (Temporal dependence)   For some $q\geq 4$ and $\chi\in(0,1)$,  $\delta_{k,q}=\O(\chi^k)$.  For some positive constant $B_\varepsilon$,   $\max_{1\leq i\leq n}\E\left|\varepsilon_i \right|^q\leq B_\varepsilon$. For all sequences $h_n\to\infty$ and $h_n<n$, $\lim_{n\to\infty}\min_{1\leq i\leq n-h_n}\E\left(\sum_{j=i}^{i+h_n}\varepsilon_j\right)^2=\infty$. 
    
 \item[(A2)] (Minimum jump magnitude and spacing) Assume that $\min_{1\leq l\leq r_n}\left|\beta(\tau_l+)-\beta (\tau_l-)\right|\geq \Delta_n\asymp n^{-\varpi}$ for some $\varpi\geq 0$ and $\min_{0\leq l\leq r_n}|\tau_{l+1}-\tau_l|\geq \gamma_n\asymp n^{ \Theta-1}$ for some $\Theta\in(0,1]$.

    \item[(A3)] (Window size) As $n\to\infty$, $m\asymp n^{\omega}\to\infty$  
  for some  $ \omega\in(0,\Theta)$. 

     \item[(A4)] The parameters specified in (A2)-(A3) satisfy  that $\Theta> 6/7 + 4\varpi/7$ and \begin{align*}
    \max\left(2/q,3-3\Theta, 2-2\Theta+2\varpi, 9-10\Theta+6\varpi\right) <  \omega < (7+4/q)\Theta-(6+4/q).
\end{align*}
\end{itemize}
    Assumption  (A1) involves a moment  and a short-range temporal dependence condition, which are commonly used in the literature. It also requires a natural nonsingularity condition that is also imposed in \cite{bonnerjee2024gaussian}. Assumption (A2)  requires that the minimum jump magnitude is bounded below by $\Delta_n$, and the minimum spacing between adjacent change points $\{\tau_l\}_{l=0}^{r_n+1}$ to be at least  $\gamma_n$. Assumption (A3) imposes a mild condition on the order of the window size $m$.    
{\color{black}The requirement on the  parameters is stipulated in Assumption (A4). 
The admissible region is non-empty. For instance, for sufficiently large \(q\), the choice $\omega={1}/{3}$ under
$\varpi=0$  and $\Theta=0.91$
satisfies all inequalities in Assumption (A4).}

\begin{theorem}\label{THM:theorem1}
    Consider Algorithm~\ref{algorithm1}, where the significance level \(\alpha\) is either fixed, or allowed to depend on \(n\) satisfying \(\alpha=\alpha_n\to0\) and  
\(
    \alpha_n \gtrsim (\log n)^{-a}
\)
for some constant \(a>0\) as $n\to\infty$.\begin{itemize}
        \item[(a)] Under Assumptions (A1)-(A3), if there is no change point, then \begin{align*}
            \P\left(\widehat r_n>0\right)=\alpha\left\{1+ \o(1)\right\}.
        \end{align*}
       \item[(b)] Under Assumptions (A1)-(A4), 
       if there are $r_n\geq 1$ jumps, then for some $C>0$, \begin{align*}
            \P\left( \left\{\widehat r_n = r_n\right\}\bigcap \max_{1\leq l\leq r_n}\left|\widehat k_l-k_l^\ast\right|\leq C \gamma_n^{-2}\Delta_n^{-2}\log n\right)\geq 1-\alpha\left\{1+ \o(1)\right\}.
        \end{align*}
    \end{itemize}
\end{theorem}

The first assertion of Theorem \ref{THM:theorem1} shows that when change points are absent, 
the BARBS procedure asymptotically controls the Type-I error at any fixed prespecified significance level $\alpha$. Moreover, if $\alpha=\alpha_n\to0$, the probability that BARBS falsely detects any change point converges to zero. 
This crucial property marks a key distinction from many existing multiple change point detection methods like \cite{killick2012optimal} and \cite{Fryzlewicz_2014}, which often fail to control false positives when no change points are present, as also illustrated in our simulation studies in Section \ref{sec:simulation}.

The second assertion of Theorem~3.1 shows that, in the presence of multiple
change points, BARBS simultaneously recovers the correct number of change
points and localizes all of them within an error of order
$\gamma_n^{-2}\Delta_n^{-2}\log n$, with asymptotic probability at least
$1-\alpha$. In particular, when $\alpha=\alpha_n\to0$, our procedure consistently estimates the number of change points and the estimated change points satisfy that 
$$\max_{1\le l\le r_n}
    \left|\widehat{k}_l-k_l^\ast\right|
    =
    \O_p\left(\gamma_n^{-2}\Delta_n^{-2}\log n\right).$$
    Comparable uniform convergence rates are obtained by \cite{Fryzlewicz_2014} under the assumption of i.i.d. Gaussian error process.  
The existing literature on binary segmentation typically imposes assumptions on the minimum spacing between consecutive change points. When all jump magnitudes are fixed constants bounded away from zero, \cite{Fryzlewicz_2014} requires \(\Theta>3/4\), and \cite{rice2022consistency} assume  \(\Theta>7/8\). While our condition in Assumption (A4) requires at least \(\Theta>9/10\), 
mainly due to the use of the multiplier bootstrap and the need to control the overall false positive probability across the stages of the recursive  procedure.

\begin{remark}
    When estimating the extreme quantile 
\(\widehat{\mathrm{crit}}_{1-\widetilde\alpha}(\mathcal T_{s:e,m})\)  in (\ref{def:crit}) 
empirically, the number \(B\) of bootstrap replications should be sufficiently large, since
  the smallest tail probability is of order \(\alpha_n\gamma_n\). Hence, a necessary requirement for stable empirical quantile estimation is
$B\alpha_n\gamma_n\to\infty$. 
\end{remark}

We now explain why the classical approach of determining the critical value of the test statistic   proposed in \citet{bai1997estimating}  generally fails in the settings of nonstationary noise under $H_0$.  
Consider a special case of our general nonstationary framework in which the
error process \(\{\varepsilon_i\}_{i\in\mathbb Z}\) is   locally stationary ($PLS(0)$ in Remark \ref{remark1}) with a
continuous long-run variance function \(\sigma^2(\cdot):=\sum_{h=-\infty}^{\infty}\Cov(\mathcal G_0(\cdot,\mathcal F_0),\mathcal G_0(\cdot,\mathcal F_h))\).
In this setting, the  F-type test statistic proposed in \citet{bai1997estimating} satisfies
\begin{align*}
    \max_{\eta n\leq k\leq (1-\eta)n}
    \frac{n^2}{k(n-k)}S_{1,k,n}^2
    \xrightarrow{d}
    \sup_{\eta\leq t\leq 1-\eta}
    \frac{\left|U(t)-tU(1)\right|^2}{t(1-t)},\quad 0<\eta<1/2,
\end{align*}
where \(U(t)\) is a mean-zero Gaussian process with covariance function
\begin{align*}
    \Cov\left\{U(t),U(t')\right\}
    =
    \int_0^{\min\{t,t'\}}\sigma^2(v)\,dv .
\end{align*}
This weak limit differs from the classical stationary limit
\begin{align*}
    \max_{\eta n\leq k\leq (1-\eta)n}
    \frac{n^2}{k(n-k)}S_{1,k,n}^2
    \xrightarrow{d}
    \sigma^{2}_\varepsilon
    \sup_{\eta\leq t\leq 1-\eta}
    \frac{\left|B(t)-tB(1)\right|^2}{t(1-t)},
\end{align*}
for some constant \(\sigma^2_\varepsilon>0\), where \(B(\cdot)\) denotes a standard
Brownian motion. 
This discrepancy highlights the inadequacy of using stationary-based asymptotic distributions to construct critical values under general nonstationary settings.

\section{Refined Estimators and Confidence Intervals} \label{sec:repartition}
In the following, we refine the initial estimators   produced by BARBS, and  further investigate the limiting distributions. 
Following \cite{bai1997estimating}, 
we introduce a second-stage refinement procedure. Conditional on the event that Algorithm \ref{algorithm1} consistently estimates the number and rough locations of the change points, we repartition the data according to the initial estimates and re-estimate each change point locally. The resulting ordered second-stage estimator is denoted by $\widetilde k_1<\widetilde k_2<\cdots<\widetilde k_{\widehat r_n}$. 
Define $\widehat k_0=0$ and $\widehat k_{\widehat r_n+1}=n$.  For $\widetilde k_l$, $1\leq l\leq \widehat r_n$,   we   locally adopt CUSUM statistics within neighborhoods around $\widehat k_l$.   More specifically,  define \begin{align}\label{DEF:widetilde k}
    \widetilde k_l=\argmax_{\widehat k_{l-1}+m\leq k\leq \widehat k_{l+1}-m}\left|\sum_{i=\widehat k_{l-1}+1}^{k}X_i-\frac{k-\widehat k_{l-1}}{\widehat k_{l+1}-\widehat k_{l-1}}\sum_{i=\widehat k_{l-1}+1}^{\widehat k_{l+1}}X_i\right|. 
\end{align}

Let $\{\varepsilon_i^\prime\}_{i\in\mathbb Z}$ be an independent copy of $\{\varepsilon_i\}_{i\in\mathbb Z}$, define \begin{align*}
    \Xi_{k^\ast}(t)= -\sum_{i=k^\ast+t+1}^{k^\ast}\varepsilon_i^\prime\mathbf{1}(t<0)+\sum_{i=k^\ast+1}^{k^\ast+t}\varepsilon_i^\prime\mathbf{1}(t>0), 
\end{align*}
and a shifting function 
\[
m_l(t)=
\begin{cases}
(\tau_{l+1}-\tau_l)/(\tau_{l+1}-\tau_{l-1}), & t<0,\\
0, & t=0,\\
(\tau_l-\tau_{l-1})/(\tau_{l+1}-\tau_{l-1}), & t>0.
\end{cases}
\]
\begin{theorem}\label{THM:asymptotic_distribution}
    Suppose that Assumptions (A1)-(A4) hold, and $\Delta_l$ are fixed nonzero constants for each $1\leq l\leq r_n$.  For any $\alpha\to 0$ satisfying the condition  in Theorem \ref{THM:theorem1},  as $n\to\infty$,  \begin{align*}
     \widetilde k_{l}-k_l^\ast\xrightarrow{d}\argmax_{t\in\mathbb Z}  \left\{-\Delta_l \Xi_{k_l^\ast}(t)-\Delta_l^2 |t|m_l(t)\right\}.
\end{align*} 
\end{theorem}
Theorem \ref{THM:asymptotic_distribution} shows that, when the jump magnitude  is fixed and nonzero, the asymptotic distribution of $\widetilde k_l$ depends on the exact distribution of the nonstationary error process and the location of the change point, which are  typically unknown.
This motivates us to consider the scenario in which the jump magnitude shrinks to zero as $n\to\infty$.

{\color{black}We next consider the limiting distribution of the refined
estimator when the jump magnitude vanishes.} Define $\mathcal W(t)=\mathcal W_1(-t)\mathbf{1}(t<0)+\mathcal W_2(t)\mathbf{1}(t\geq 0)$, 
in which $\{\mathcal W_1(t)\}_{t\geq 0}$ and $\{\mathcal W_2(t)\}_{t\geq 0}$ are independent Wiener processes. Unlike the preceding results, the following
Theorem~\ref{THM:asymptotic_distribution2} requires a piecewise locally stationary structure of the error process with the long-run variance
continuous at the true change point $k_l^\ast$, and thus does not retain the full generality of the genuinely nonstationary framework considered above. Under
this additional condition, the limiting distribution of the refined change point estimator $\widetilde k_l$ nevertheless becomes
simpler, depending only on the long-run variance at the change point and the
relative spacings between adjacent change points, rather than on the joint
distribution of the error process.
\begin{theorem}\label{THM:asymptotic_distribution2}
  For some $d\geq 0$, suppose that the error process $\{\varepsilon_i\}_{i\in\mathbb Z}$ is   $PLS(d)$  with a piecewise continuous long-run variance function $\sigma^2(\cdot)$,   whose discontinuity points do not coincide with the change points $\{\tau_l\}_{l=1}^{r_n}$.   Under Assumptions (A1)-(A4),  $|\Delta_l|\to 0$, and $\Delta_l^2\min\{ k_{l+1}^\ast-k_{l }^\ast ,k_{l }^\ast-k_{l-1 }^\ast\}\to\infty$   
  for each $1\leq l\leq r_n$, for any $\alpha\to 0$ satisfying the condition  in Theorem \ref{THM:theorem1},   as $n\to\infty$,  \begin{align*}
     \frac{\Delta_l^2}{\sigma^2(\tau_l)}\left(\widetilde k_{l}-k_l^\ast\right)\xrightarrow{d}\argmax_{t\in\mathbb R}  \left\{\mathcal W(t)-  |t|m_l(t)\right\}. 
\end{align*}  
\end{theorem}
Theorem  \ref{THM:asymptotic_distribution2}  indicates that an asymptotic $(1-\alpha_0)$-confidence interval for the $l$-th change point $k_l^\ast$ is given by \begin{align*}
    \left(\widetilde k_l-\frac{\widehat Q_{l,1-\alpha_0/2}\widehat\sigma^2_l}{\widehat\Delta_l^2},\widetilde k_l-\frac{\widehat Q_{l, \alpha_0/2}\widehat\sigma^2_l}{\widehat\Delta_l^2}\right),
\end{align*}
where $\widehat Q_{l,\alpha}$ is the $\alpha$-quantile of the random variable $\argmax_{t\in\mathbb R}  \left\{\mathcal W(t)-  |t|\widehat m_l(t)\right\}$, where the shifting function $m_l(t)$ is estimated by 
\[
\widehat m_l(t)=
\begin{cases}
(\widehat k_{l+1}-\widetilde k_l)/(\widehat k_{l+1}-\widehat k_{l-1}), & t<0,\\
0, & t=0,\\
(\widetilde k_l-\widehat k_{l-1})/(\widehat k_{l+1}-\widehat k_{l-1}), & t>0,
\end{cases}
\] and \begin{align*}
    \widehat\Delta_l = \frac{1}{\widehat k_{l+1}-\widetilde k_l}\sum_{i=\widetilde k_l+1}^{\widehat k_{l+1}}X_i - \frac{1}{\widetilde k_l-\widehat k_{l-1}}\sum_{i=\widehat k_{l-1}+1}^{\widetilde k_l}X_i.
\end{align*}
Besides, $\widehat{\sigma}^2_l$ is any consistent estimator of $ \sigma^2(\tau_l)$. In practice, one possible implementation of $\widehat{\sigma}^2_l$ is to first detrend the  original time series according to the change points identified by BARBS, and then apply the method of \cite{bai2024difference} with all covariates set to one.

{\color{black}The following theorem further provides a theoretical guarantee for the faster uniform convergence rate of the refined change point estimators $\widetilde k_l$, $1\leq l\leq \widehat r_n$. 
\begin{theorem} 
\label{THM:uniform_refined_rate}
  Suppose that Assumptions (A1)-(A4) hold.  For any $\alpha\to 0$ satisfying the condition  in Theorem \ref{THM:theorem1},  as $n\to\infty$,  
\[
    \max_{1\le l\le   r_n}
    \Delta_l^2|\widetilde k_l-k_l^\ast|
    =
    \O_p\left(\max\left\{ \left(\sum_{l=1}^{r_n}|\Delta_l|^{q-2}\right)^{1/(q-1)},1 \right\} \log n \right).
\]
\end{theorem}
It is noted that Theorem~\ref{THM:uniform_refined_rate} does not distinguish between fixed and vanishing jump magnitudes, and thus provides a unified uniform bound for the refined estimators. When all jump magnitudes are fixed and  bounded away from zero, the uniform convergence rate depends only on the number of change points and the moment   \(q\) of the nonstationary error process, i.e., 
\[
    \max_{1\le l\le   r_n}
    \left|\widetilde k_l-k_l^\ast\right|
    =
    \O_p\left(
    \max\left\{r_n^{1/(q-1)},1\right\}\log n
    \right).
\]
In fact, Corollary~3.5 of \cite{wu2024multiscale} gives a  concrete example which indicates that such a rate cannot be improved up to some logarithmic factor.  
When all jump magnitudes are of the same order and shrink to zero, say \(|\Delta_l|\asymp \Delta_n\) with $\Delta_n^{q-2} r_n \to 0$, Theorem~\ref{THM:uniform_refined_rate} yields that 
\[
    \max_{1\le l\le   r_n}
    \left|\widetilde k_l-k_l^\ast\right|
    =
    \O_p\left(\Delta_n^{-2}\log n\right),
\]
which agrees with the optimal localization rate up to a logarithmic factor.  
}

\section{Numerical Simulations}\label{sec:simulation}

\subsection{Tuning parameter  selection}
For BARBS, we use \( B = 2000 \) bootstrap replications to compute the critical values, and select the block size \( m \) via the plug-in rule described in \cite{wu2024frequency}. 
{\color{black}Specifically,
we  note that $L_n:=\E (\sum_{i=1}^{n}\varepsilon_i)^2/n$ is approximated by { \begin{align*}
 				\widehat L_{n,m} = \sum_{j=m}^{n-m}Y_{j,m }^2/(n-2m+1),
 			\end{align*}}
 		where $Y_{j,m}$, $m\leq j\leq n-m$, is defined in (\ref{def:Yj}).
        From \cite{wu2024frequency}, 
 		  	$\Var(\widehat L_{n,m})= m\left\{C_{1n}^2+\o(1)\right\} /n$  and  ${\text{bias}}(\widehat L_{n,m})= \{C_{2n}+\o(1)\}/m$, for some $C_{1n}$ and $C_{2n}$ depending on the data $\{X_i\}_{i=1}^{n}$. Let $m'=[n^{1/3}]$ and $l_n=[n^{1/6}]$. We estimate  $C_{1n}$ and $C_{2n}$ by \begin{align*}
 		  	    &\widehat C_{1n}^2=\sum_{i=2m'}^{n-2m'} H_{im'}^2/\{(n-4m'+1)m'\},\quad H_{im'}=\left(\sum_{j=i-m'+1}^{i}Y_{j,m'}^2-\sum_{j=i+1}^{i+m'}Y_{j,m'}^2\right)/\sqrt{2m'}.\\
                &\widehat C_{2n}=3/(n-3l_n+1)\sum_{j=1}^{n-3l_n+1}s_{j,l_n}(s_{j+l_n,l_n}-s_{j+2l_n,l_n}),\quad s_{j,l_n}=\sum_{i=j}^{j+l_n-1}X_i.
 		  	\end{align*} 
Finally, we select $m=\max\left\{1,\left[\left(2\widehat C_{2n}^2 n/\widehat C_{1n}^2 \right)^{1/3}\right]\right\}$ to minimize the  MSE of $\widehat L_{n,m}$. 
}

\subsection{Benchmarks}
We compare BARBS with several representative modern change point detection techniques, including BAI \citep{bai1997estimating}, WBS \citep{Fryzlewicz_2014}, PELT \citep{killick2012optimal}, HSMUCE \citep{pein2017heterogeneous}, and MOSUM \citep{eichinger2018mosum}. 
BAI is a binary segmentation method that determines its critical values using a test statistic involving the maximum of a normalized Brownian bridge. We simulate  1000 Brownian bridge samples of length 1000 to compute the critical values, and modify the significance level of each test in BAI  following BARBS described in Section~\ref{sec:bsprocedure}. The hyperparameter \( \eta \), which determines the proportion of data discarded at the beginning and end of the time series, is set to \( \eta = 0.01 \). 
WBS performs change point detection on a collection of randomly drawn intervals from the time series. We use the \texttt{wbs} package for implementation and apply the Modified Bayesian Information Criterion to determine the number of change points. 
PELT is a penalized cost-based method that uses pruning to achieve efficient   optimization. It is implemented via the \texttt{changepoint} package, using the default penalty function (Modified Bayesian Information Criterion). 
HSMUCE is a multiscale testing procedure whose critical values are approximated via Monte Carlo simulations. We use the \texttt{stepR} package with 1000 Monte Carlo replicates to estimate the critical values. Finally, 
MOSUM is a moving sum change point detection method that allows for multiple bandwidths in testing. 
We implement MOSUM using the \texttt{mosum} package and apply the multiscale MOSUM procedure with localized pruning.

\subsection{Data generating processes}\label{sec:DGP}

We consider the following models for the error process $\{\varepsilon_i\}_{i=1}^n$: independent and identically distributed (IID), autoregressive (AR), piecewise stationary (PS), locally stationary (LS), piecewise locally stationary (PLS), nonlinear (NL), and time-varying moving average (TVMA) processes.

\begin{itemize}
    \item[(IID)] $\varepsilon_i=\eta_i$ where $\eta_i\sim\mathcal{N}(0,1)$.
    \item[(AR)]
    $\varepsilon_i= \varepsilon_{i-1}/3+\eta_i$ where $\eta_i\sim\mathcal{N}(0,1)$. 
    \item[(PS)] $\varepsilon_i=\eta_i$ where $\eta_i$ are $\mathcal{N}(0,10^2)$ for $i\leq \lfloor n/4\rfloor$, and $\mathcal{N}(0,1)$ otherwise.
    \item[(LS)]$\varepsilon_i=a_i(\eta_{i-1}+\eta_{i-2})+\eta_i$, where $\eta_i$'s are centered and standardized to unit variance draws from a Binomial(10,0.5) distribution, and $a_i=\cos \left( {2i\pi}/{n} \right)+1$. 
    \item[(PLS)]
    $\varepsilon_i= \varepsilon_{i-1}/3+a_i\eta_i$, where $\eta_i$ follows the centered    exponential distribution with scaling parameter $1$, i.e., $\eta_i\sim \text{Exp}(1)-1$ and $a_i=\cos\left( {2\pi i}/{n}\right)+2 + 5 \mathbf{1}_{i> \lfloor n/2 \rfloor} $.
    \item[(NL)] The errors follow a Markov switching model with time-varying transition probabilities
\begin{align*}
	\varepsilon_i&=\begin{cases}
	\varepsilon_{i-1}/2+\eta_i & S_i=0, \\
		- \varepsilon_{i-1}/2+\eta_i & S_i=1.\\
	\end{cases}
\end{align*}
with $\eta_i\sim\text{Unif}(- {1}/{\sqrt{12}}, {1}/{\sqrt{12}}), $ and where at initialization $S_0=0$ and the probability of switching between states at index $i$ is $p_{i,01}=p_{i,10}=0.05\{1- {(i-1)}/{(n-1)}\}$. This process is similar to that in \cite{ding2023autoregressive}.
\item[(TVMA)] Let $\{\eta_i\}_{i\in\mathbb Z}$ be a  sequence of i.i.d. Gaussian random variables. Define $\varepsilon_i=\sum_{l=0}^{L}\rho^l a_{il}\eta_{i-l}$. 
For each $i$, the random coefficients $\{a_{il}\}_{l=0}^{L}$ are i.i.d.\ conditional on $A_i$, with $a_{il}{\sim} \text{Unif}(-A_i,A_i)$, where   $A_i  \overset{\mathrm{i.i.d.}}{\sim}  \text{Unif}(2,4)$ when $ i<n/3$ or $i>2n/3$, and $A_{i}  \overset{\mathrm{i.i.d.}}{\sim} \text{Unif}(0,2)$ when $n/3\leq i\leq 2n/3$. The  parameters $L=30$ and $\rho=0.9$.  

\end{itemize}
 
\subsection{Detection accuracy under alternatives}\label{sec:simulation1}

To evaluate the performance of BARBS for detecting multiple change points,   we consider several signal configurations under the alternative hypothesis. In all cases, SNR refers to the absolute value of jump magnitude relative to the sample standard deviation of the simulated process. 

\begin{itemize}

	\item[(1)] A single change point   occurs at $ {n}/{2}$ with ${\text {SNR}}=1$.
	\item[(2)] Two change points  occur at $\lfloor  {n}/{3} \rfloor$ and $\lfloor  {2n}/3\rfloor$ with ${\text {SNR}}=1$; the sign of each change point is chosen independently to be positive or negative with equal probability. 
	 \item [(3)]  Four change points of various sizes (${\text {SNR}}$ equals 0.5, 1.0, 1.5, 2.0) occur at  $\lfloor  {n}/{5} \rfloor$, $\lfloor  {2n}/{5} \rfloor$, $\lfloor  {3n}/{5} \rfloor$, $\lfloor  {4n}/{5} \rfloor$. The sign of each change point is positive or negative with equal probability, and the order in which the different sizes of change points occur is random.

\end{itemize}

Across all simulations, we conducted 500 independent experiments and computed summary statistics as outlined below.  
An estimated change point $\widehat k_l$ is said to be correctly detected if there exists a true change point $k_l^\ast$ such that 
\(
    |\widehat k_l-k_l^\ast|\le [1+\log n].
\)
We report the true number of change points $r_n$, the average number ($\hat{r}_n$) of detected change points, the proportion ($\hat{r}_n^\ast/r_n$) of correctly identified change points, the percentage  ($s$) of replications in which all change points were correctly detected, and the percentage ($s^\ast$) of replications in which all change points were detected with no false positives. In particular, \(s^\ast\)  is the most   informative as it captures both detection power and false positive control. We also report the mean absolute deviation (MAD) between the estimated and true locations for correctly identified change points.
Furthermore, we include the average Adjusted Rand Index (ARI) \citep{hubert1985comparing}, a widely used metric for evaluating clustering and segmentation performance. The ARI is bounded above by 1, with 1 indicating perfect agreement; values near 0 correspond to agreement expected under random partitioning.

Based on the results presented in  Tables \ref{tab:scenario1-PLS}-\ref{tab:scenario3-PLS} and Tables S.1-S.18 in the Supplementary Materials, we summarize the following findings. When the error is i.i.d., all methods perform reasonably well. However, when the error exhibits complex dependence structures, methods such as WBS, PELT, and MOSUM tend to produce a large number of false positives. This is evident in the  PS, LS, PLS, NL, and TVMA settings, where these methods, or some of them, identify far more change points than truly exist. 
Moreover, HSMUCE also demonstrates relatively high false positive rates in the LS settings of Scenario 1 and 2, while it tends to severely underestimate the number of change points in Scenario 3 under the IID, AR, NL, and TVMA settings. In addition, among the competing methods, BAI exhibits the most similar performance to BARBS. However, under the PS, LS, PLS, and NL settings across different scenarios, BAI tends to detect more incorrect change points than BARBS, resulting in significantly lower values of \(s^\ast\). This indicates that BAI may not   control false positives well under nonstationary time series. 
Overall, BARBS provides more stable false positive control across different nonstationary settings, while maintaining competitive detection accuracy.

\begin{table}[h!]
  \centering
  \caption{Scenario 1 with PLS}
  \resizebox{\textwidth}{!}{
    \begin{tabular}{ccccccccccccccc}
    \toprule
          & \multicolumn{7}{c}{$n=500$}                             & \multicolumn{7}{c}{$n=1000$} \\
\cmidrule{2-15}          & $r_n$     & $\hat r_n$  & $\hat r_n^\ast/r_n$ & $s$     & $s^\ast$ & MAD   & ARI   & $r_n$     & $\hat r_n$  & $\hat r_n^\ast/r_n$ & $s$     & $s^\ast$ & MAD   & ARI \\
    \midrule
    BARBS   & 1     & 1.472  & 0.834  & 0.834  & 0.578  & 1.758  & 0.903  & 1     & 1.326  & 0.846  & 0.846  & 0.654  & 1.712  & 0.939  \\
    BAI   & 1     & 2.068  & 0.930  & 0.930  & 0.440  & 1.594  & 0.899  & 1     & 1.970  & 0.918  & 0.918  & 0.482  & 1.638  & 0.923  \\
    WBS   & 1     & 16.546  & 0.854  & 0.854  & 0.004  & 1.810  & 0.567  & 1     & 21.118  & 0.870  & 0.870  & 0.002  & 1.687  & 0.567  \\
    PELT  & 1     & 116.692  & 0.990  & 0.990  & 0.000  & 0.909  & 0.129  & 1     & 219.942  & 0.996  & 0.996  & 0.000  & 0.898  & 0.096  \\
    HSMUCE & 1     & 1.380  & 0.942  & 0.942  & 0.618  & 1.571  & 0.901  & 1     & 1.848  & 0.882  & 0.882  & 0.340  & 1.333  & 0.825  \\
    MOSUM & 1     & 2.188  & 0.814  & 0.814  & 0.348  & 1.799  & 0.830  & 1     & 3.244  & 0.816  & 0.816  & 0.198  & 1.858  & 0.795  \\
    \bottomrule
    \end{tabular}%
    }
  \label{tab:scenario1-PLS}%
\end{table}%

\begin{table}[h!]
  \centering
  \caption{Scenario 2 with PLS}
  \resizebox{\textwidth}{!}{
    \begin{tabular}{ccccccccccccccc}
    \toprule
          & \multicolumn{7}{c}{$n=500$}                             & \multicolumn{7}{c}{$n=1000$} \\
\cmidrule{2-15}          & $r_n$     & $\hat r_n$  & $\hat r_n^\ast/r_n$ & $s$     & $s^\ast$ & MAD   & ARI   & $r_n$     & $\hat r_n$  & $\hat r_n^\ast/r_n$ & $s$     & $s^\ast$ & MAD   & ARI \\
    \midrule
    BARBS   & 2     & 2.514  & 0.816  & 0.634  & 0.410  & 1.044  & 0.902  & 2     & 2.380  & 0.800  & 0.602  & 0.434  & 1.037  & 0.937  \\
    BAI   & 2     & 3.020  & 0.827  & 0.656  & 0.312  & 1.060  & 0.895  & 2     & 3.034  & 0.819  & 0.638  & 0.326  & 1.074  & 0.929  \\
    WBS   & 2     & 17.410  & 0.901  & 0.802  & 0.006  & 1.086  & 0.648  & 2     & 22.390  & 0.872  & 0.744  & 0.002  & 1.070  & 0.637  \\
    PELT  & 2     & 117.694  & 1.000  & 1.000  & 0.000  & 0.388  & 0.181  & 2     & 220.908  & 1.000  & 1.000  & 0.000  & 0.382  & 0.146  \\
    HSMUCE & 2     & 2.226  & 0.627  & 0.262  & 0.198  & 0.746  & 0.792  & 2     & 2.708  & 0.691  & 0.386  & 0.194  & 0.727  & 0.838  \\
    MOSUM & 2     & 3.072  & 0.802  & 0.614  & 0.266  & 0.959  & 0.866  & 2     & 4.162  & 0.805  & 0.620  & 0.144  & 0.977  & 0.856  \\
    \bottomrule
    \end{tabular}%
    }
  \label{tab:scenario2-PLS}%
\end{table}%
 
\begin{table}[h!]
  \centering
   \caption{Scenario 3 with PLS}
  \resizebox{\textwidth}{!}{
    \begin{tabular}{ccccccccccccccc}
    \toprule
          & \multicolumn{7}{c}{$n=500$}                             & \multicolumn{7}{c}{$n=1000$} \\
\cmidrule{2-15}          & $r_n$     & $\hat r_n$  & $\hat r_n^\ast/r_n$ & $s$     & $s^\ast$ & MAD   & ARI   & $r_n$     & $\hat r_n$  & $\hat r_n^\ast/r_n$ & $s$     & $s^\ast$ & MAD   & ARI \\
    \midrule
    BARBS   & 4     & 5.104  & 0.804  & 0.364  & 0.150  & 1.552  & 0.836  & 4     & 4.988  & 0.815  & 0.402  & 0.200  & 1.677  & 0.890  \\
    BAI   & 4     & 5.786  & 0.811  & 0.382  & 0.078  & 1.405  & 0.832  & 4     & 5.912  & 0.825  & 0.420  & 0.140  & 1.629  & 0.883  \\
    WBS   & 4     & 18.710  & 0.865  & 0.584  & 0.002  & 1.859  & 0.696  & 4     & 23.892  & 0.887  & 0.606  & 0.000  & 2.112  & 0.714  \\
    PELT  & 4     & 118.696  & 1.000  & 1.000  & 0.000  & 0.462  & 0.297  & 4     & 221.878  & 1.000  & 0.998  & 0.000  & 0.539  & 0.231  \\
    HSMUCE & 4     & 3.548  & 0.666  & 0.106  & 0.104  & 1.629  & 0.773  & 4     & 4.160  & 0.710  & 0.194  & 0.124  & 1.268  & 0.840  \\
    MOSUM & 4     & 4.498  & 0.753  & 0.288  & 0.102  & 1.494  & 0.804  & 4     & 5.808  & 0.803  & 0.388  & 0.104  & 1.773  & 0.859  \\
    \bottomrule
    \end{tabular}%
    }
  \label{tab:scenario3-PLS}%
\end{table}%

\subsection{Type-I error under the null}
To evaluate the performance of different change point detection methods when no change points are present in the time series, we now
consider that the trend function $\beta(\cdot)\equiv 0$, and the DGPs are described in Section \ref{sec:DGP}.  

\begin{figure}[h!]
    \centering
    \includegraphics[width=0.48\linewidth]{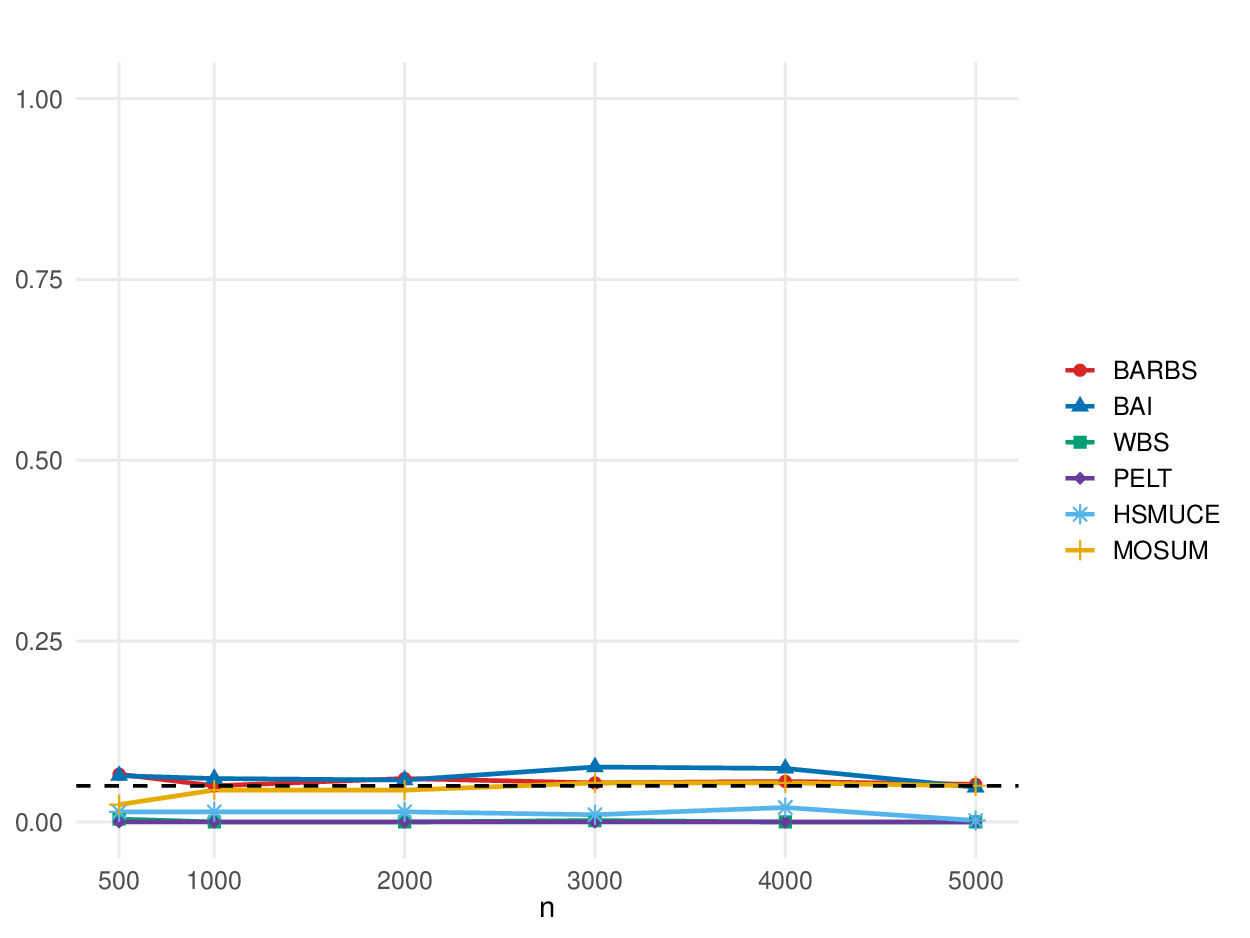}
    \includegraphics[width=0.48\linewidth]{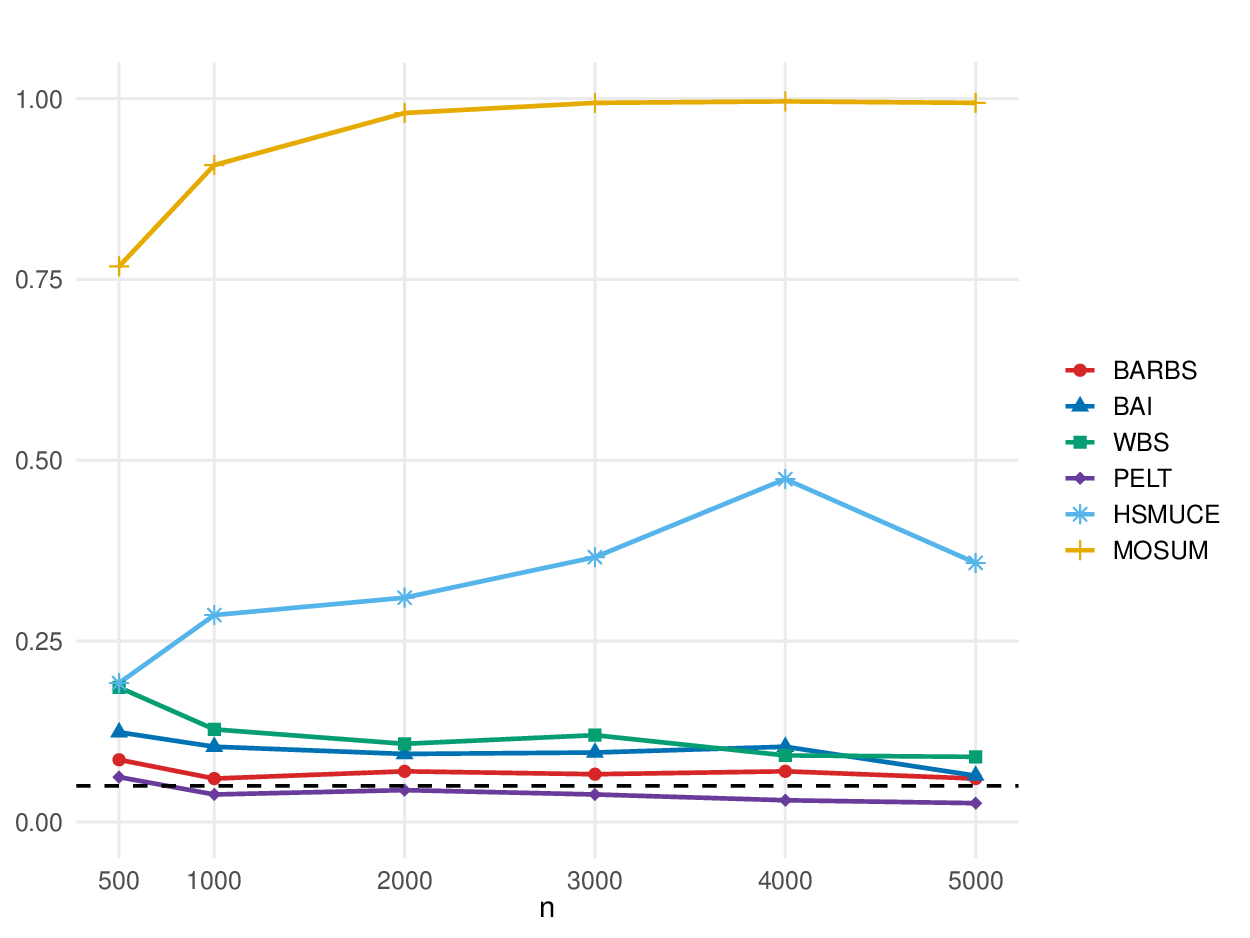}
    \includegraphics[width=0.48\linewidth]{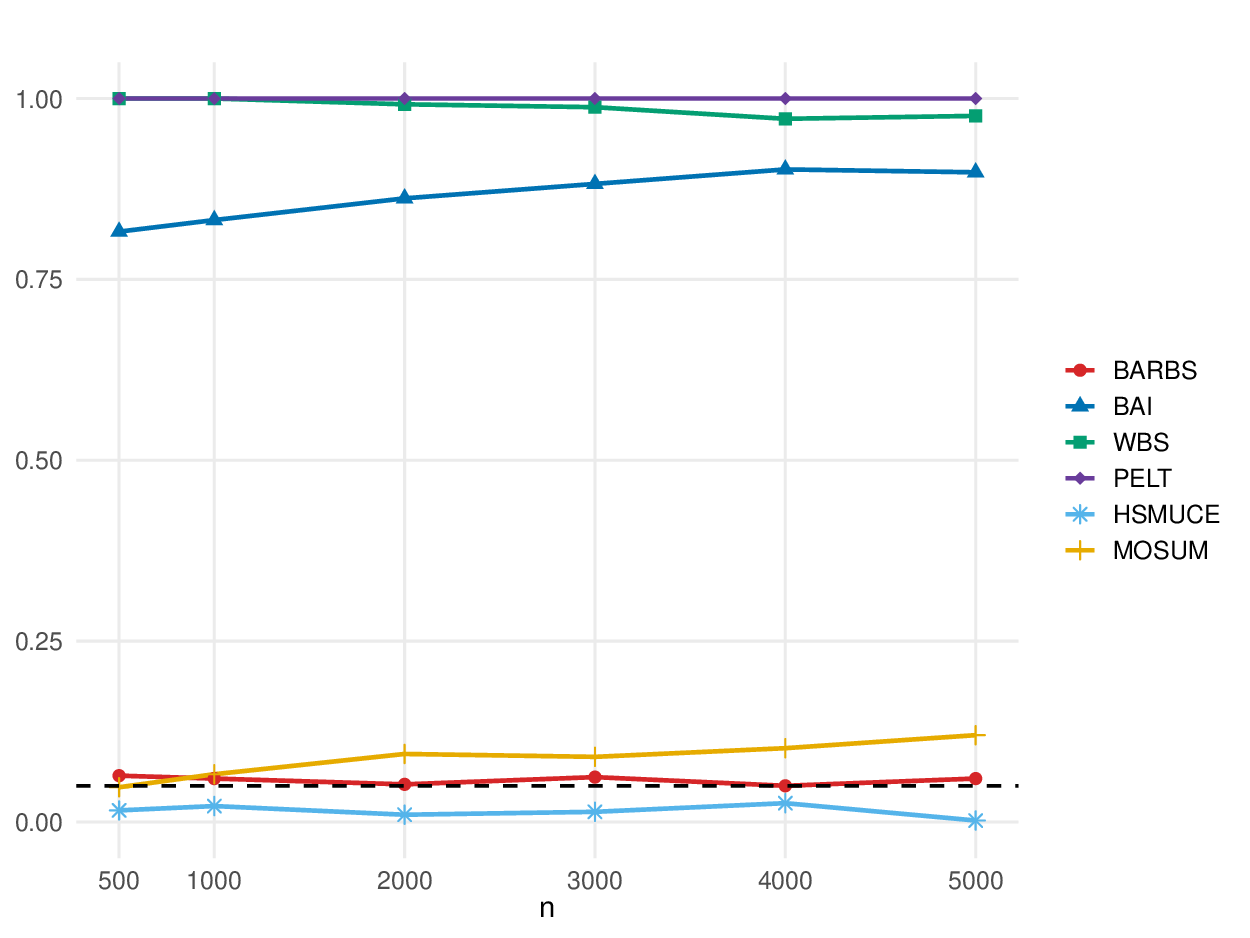}
    \includegraphics[width=0.48\linewidth]{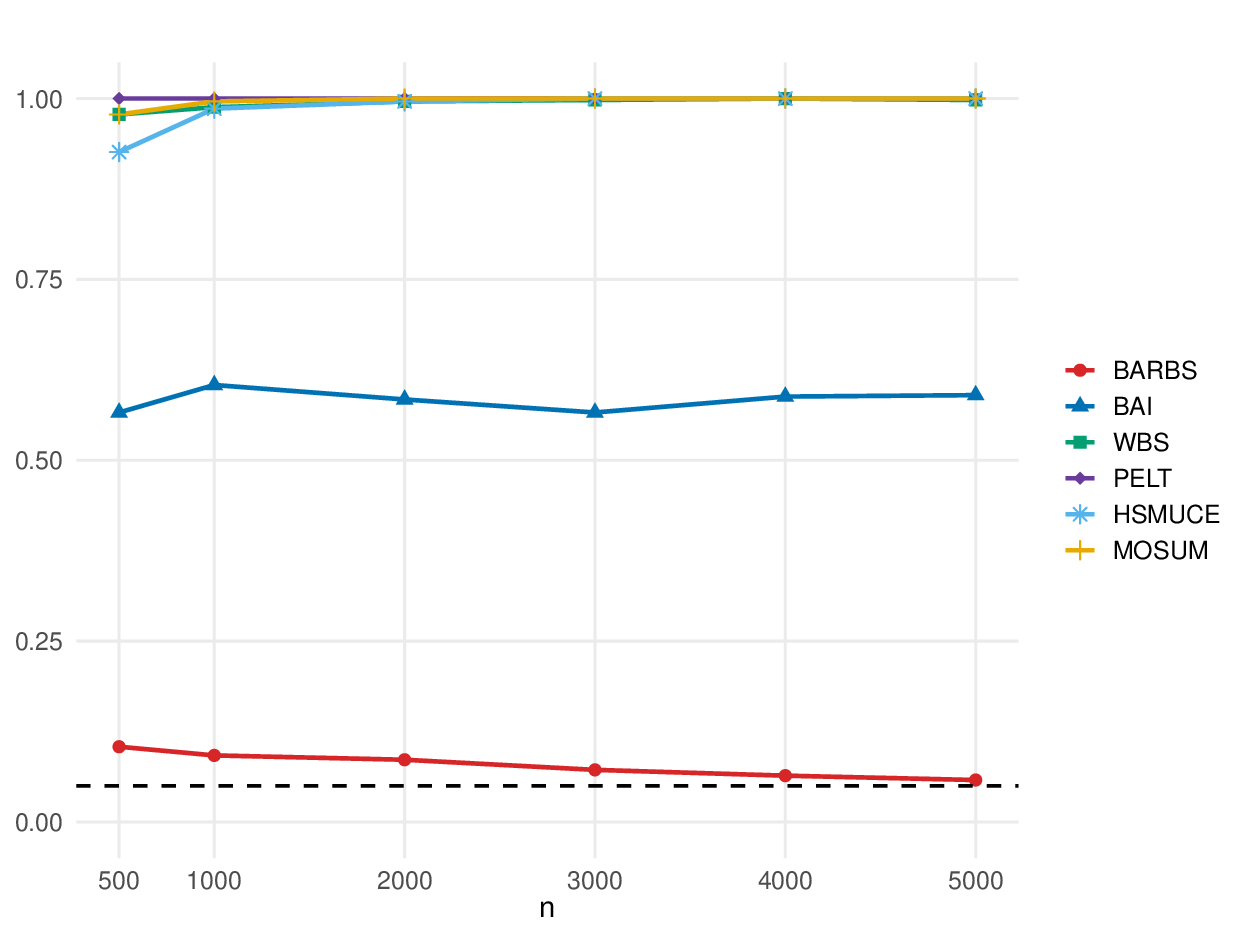}
    \caption{Empirical Type-I error of different multiple change point detection procedures in the absence of change points. Top-left: IID, Top-right: AR, Bottom-left: PS, Bottom-right: LS. }
    
    \label{fig:simu1}
\end{figure}

\begin{figure}[h!]
    \centering
    
\includegraphics[width=0.48\linewidth]{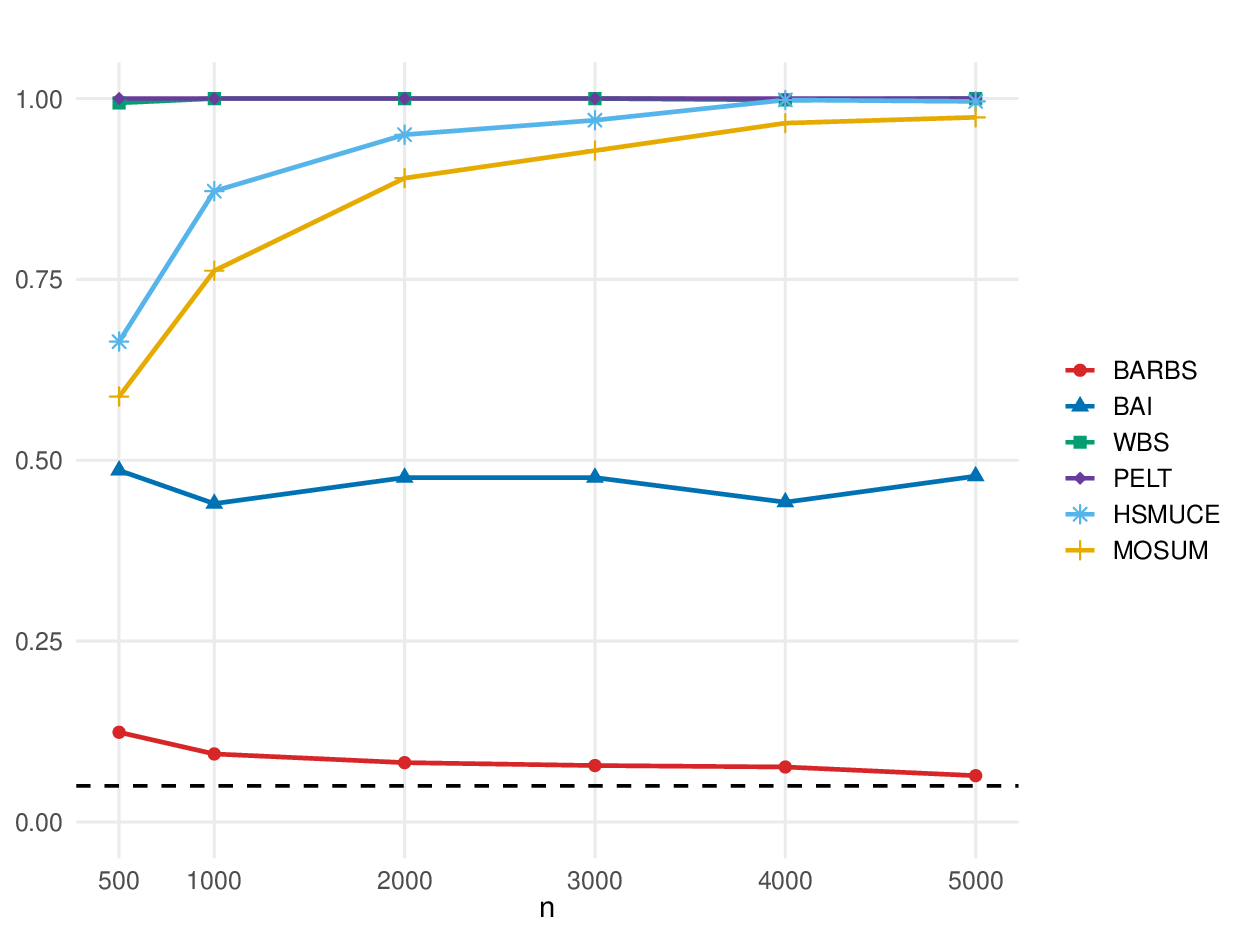}

    \includegraphics[width=0.48\linewidth]{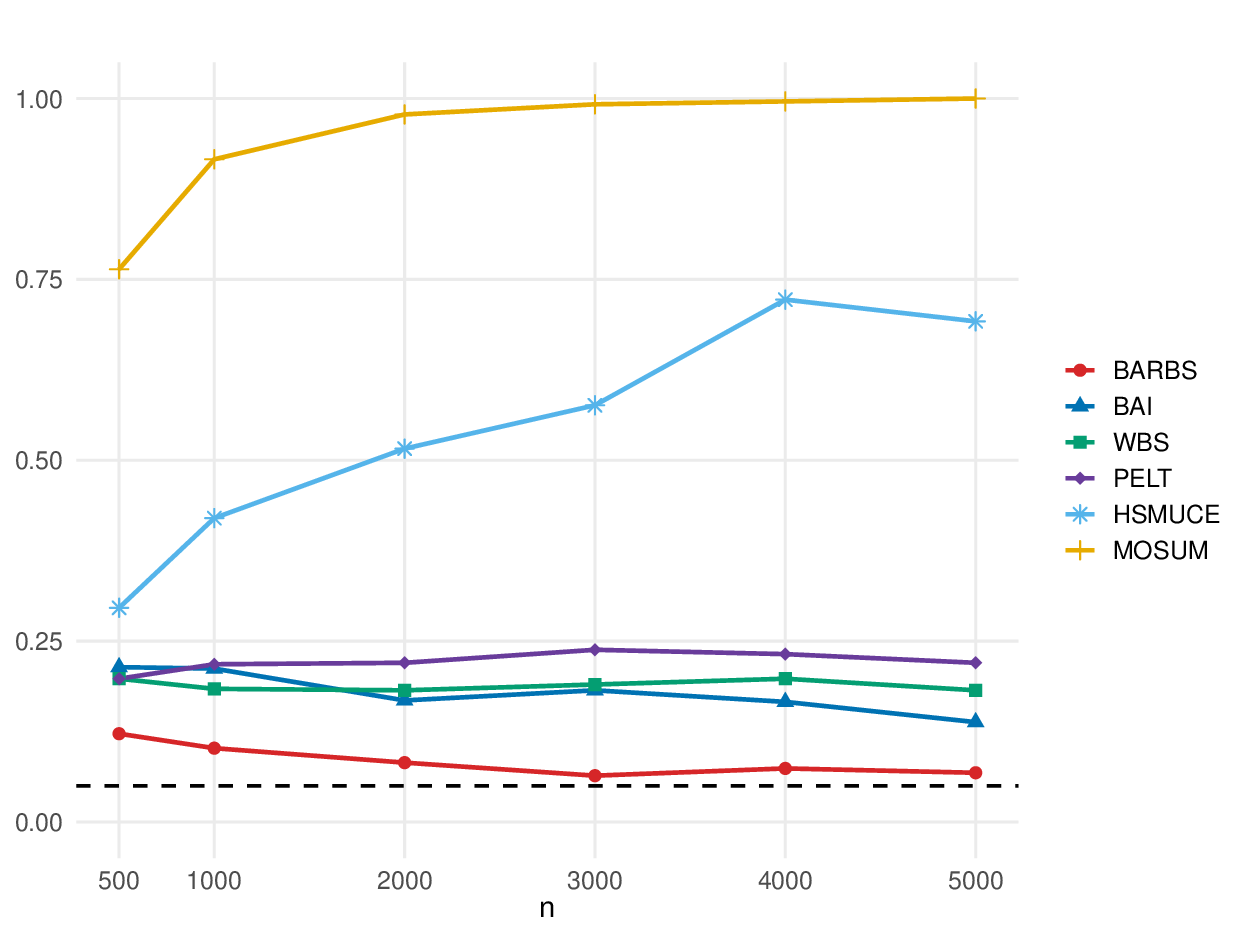}
    \includegraphics[width=0.48\linewidth]{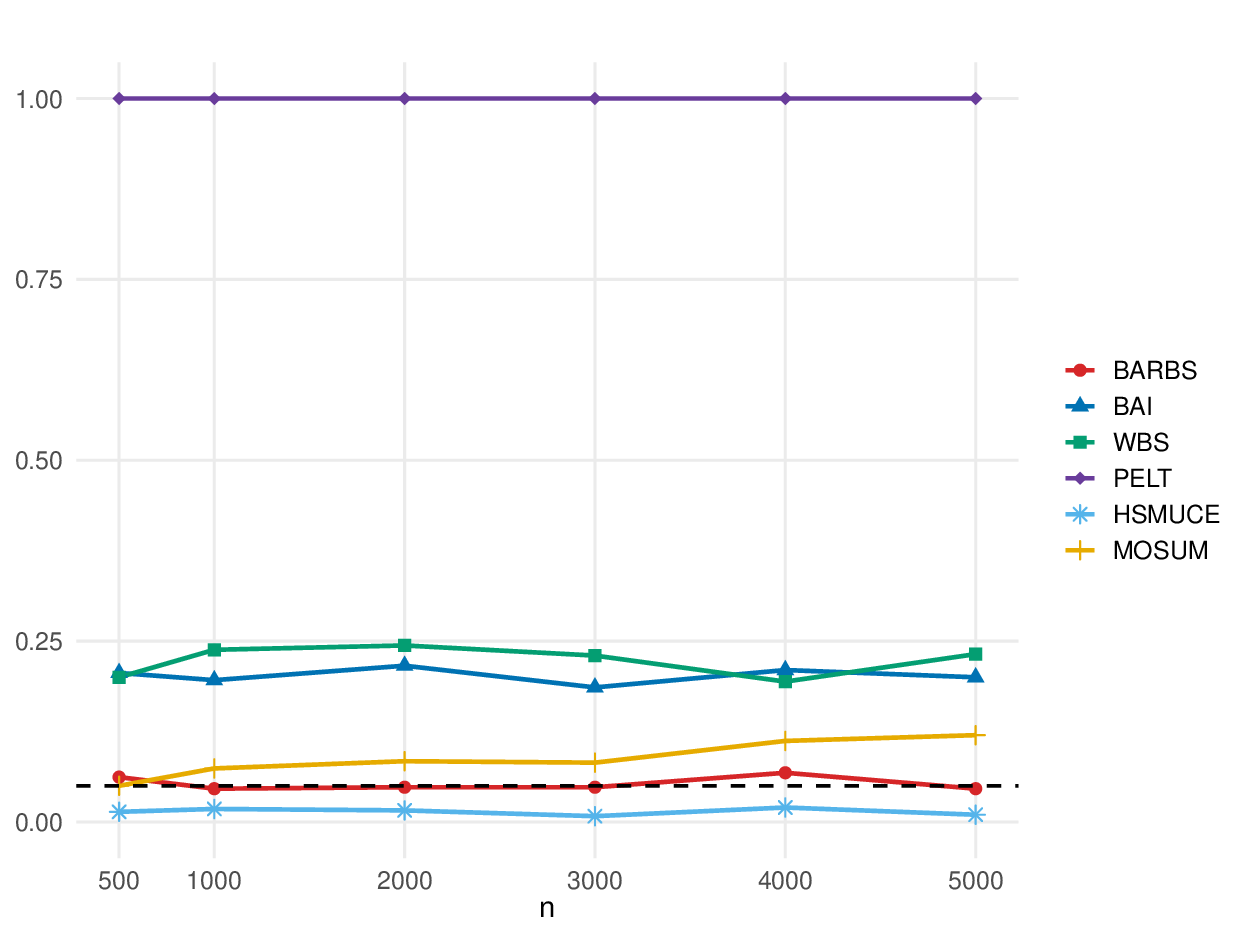}
   \caption{Empirical Type-I error of different multiple change point detection procedures in the absence of change points. Top: PLS,   Bottom-left: NL, Bottom-right: TVMA. }
      
    \label{fig:simu2}
\end{figure}

The results in Figures \ref{fig:simu1}-\ref{fig:simu2} demonstrate that BARBS effectively controls the Type I error near the nominal level $\alpha=0.05$ across a wide range of DGPs as the sample size increases. In contrast, the competing methods only maintain proper Type I error control under IID or stationary AR settings. When the noise structure deviates from the stationarity assumption, all other methods, including BAI, tend to produce elevated false positive rates.

\subsection{Confidence intervals of change points}
We conduct numerical experiments to assess the finite-sample performance of the proposed 
confidence intervals.  To this end, we examine different types of time series models: 
 (IID), (AR), (LS), (PS), (PLS).
Two signal scenarios are examined: (a) a single change point located at $\lfloor n/3\rfloor$ with SNR $=2$; (b) two change points positioned at $\lfloor n/3 \rfloor$ and $\lfloor  n/2 \rfloor$, both with SNR $ = 2$. In the second scenario, each change point is assigned a positive or negative sign with equal probability. To  compare  with existing methods, we additionally implement two alternative confidence intervals: one based on Theorem 2.8.2 of \cite{csorgo1997limit}, denoted by CH, and the other based on Theorem 2.3.3 of \cite{horvath2024change}, denoted by HR.
For a fair comparison, all confidence intervals are based on the same refined estimator  
$\widetilde k_l$, while differing  in the limit-distribution calibration.
For our proposed confidence intervals (denoted by BARBS), the local long-run variance is estimated using the \textsf{R} package \texttt{mlrv}, while
for HR, we employ the \textsf{R} package \texttt{cointReg} to obtain the global long-run variance estimator.
Coverage rates and average lengths are computed conditional on successful detection of the corresponding change point.

We fix the significance level at $\alpha = 0.01$ in Algorithm~\ref{algorithm1} for the detection step.  The results in Tables \ref{tab:CI1}-\ref{tab:CI2} show that the empirical coverage rate of the proposed $95\%$ confidence intervals is   close to the nominal confidence level when the noise process is locally stationary in a neighborhood of the change point.  In the three settings where stationarity is violated, namely LS, PS, and PLS, our confidence intervals are narrower than those of the competing methods, since the latter rely on global long-run variance or variance estimates. Moreover, it is worth noting that,  in the deliberately constructed scenario (b) under the PLS setting the second change point coincides with a break in the error structure. This violates the long-run variance continuity condition assumed in Theorem \ref{THM:asymptotic_distribution2}, leading to a reduced empirical coverage rate for the proposed confidence intervals in this case. 

\begin{table}[htbp]
  \centering
  \caption{Empirical coverage rate (CR) and average length (AL) of the proposed $95\%$ confidence intervals for scenario (a).}
  \resizebox{0.8\textwidth}{!}{
    \begin{tabular}{ccccccccccccc}
    \toprule
    $n$     & \multicolumn{6}{c}{500}                       & \multicolumn{6}{c}{1000} \\
    \midrule
          & \multicolumn{2}{c}{BARBS} & \multicolumn{2}{c}{HR} & \multicolumn{2}{c}{CH} & \multicolumn{2}{c}{BARBS} & \multicolumn{2}{c}{HR} & \multicolumn{2}{c}{CH} \\
\cmidrule{2-13}          & CR    & AL    & CR    & AL    & CR    & AL    & CR    & AL    & CR    & AL    & CR    & AL \\
\cmidrule{2-13}    IID   & 0.948  & 7.9   & 0.944  & 7.6   & 0.956  & 8.0   & 0.946  & 7.9   & 0.946  & 7.8   & 0.952  & 7.9  \\
    AR    & 0.948  & 20.0  & 0.930  & 18.1  & 0.814  & 8.1   & 0.960  & 20.0  & 0.944  & 18.9  & 0.848  & 8.0  \\
    LS    & 0.950  & 5.1   & 0.988  & 18.6  & 0.984  & 7.8   & 0.954  & 5.1   & 0.992  & 19.5  & 0.976  & 7.9  \\
    PS    & 1.000  & 1.8   & 0.996  & 7.3   & 1.000  & 8.0   & 1.000  & 1.8   & 1.000  & 7.6   & 1.000  & 8.0  \\
    PLS   & 0.984  & 1.3   & 0.992  & 13.6  & 1.000  & 7.9   & 0.980  & 1.2   & 0.998  & 14.3  & 1.000  & 7.9  \\
    \bottomrule
    \end{tabular}%
    }
  \label{tab:CI1}%
\end{table}%

\begin{table}[htbp]
  \centering
  \caption{Empirical coverage rate (CR) and average length (AL) of the proposed $95\%$ confidence intervals for scenario (b). }
  \resizebox{ 0.8\textwidth}{!}{
    \begin{tabular}{ccccccccccccc}
    \toprule
    $n$     & \multicolumn{6}{c}{500}                       & \multicolumn{6}{c}{1000} \\
    \midrule
          & \multicolumn{2}{c}{BARBS} & \multicolumn{2}{c}{HR} & \multicolumn{2}{c}{CH} & \multicolumn{2}{c}{BARBS} & \multicolumn{2}{c}{HR} & \multicolumn{2}{c}{CH} \\
    \midrule
    $k_1$    & CR    & AL    & CR    & AL    & CR    & AL    & CR    & AL    & CR    & AL    & CR    & AL \\
    \midrule
    IID   & 0.956  & 11.0  & 0.958  & 10.3  & 0.958  & 10.0  & 0.966  & 8.1   & 0.964  & 8.1   & 0.964  & 8.1  \\
    AR    & 0.944  & 23.0  & 0.932  & 20.9  & 0.810  & 12.1  & 0.966  & 24.9  & 0.952  & 24.1  & 0.852  & 9.9  \\
    LS    & 0.942  & 7.0   & 0.980  & 23.5  & 0.970  & 10.7  & 0.944  & 6.7   & 0.986  & 25.7  & 0.974  & 10.6  \\
    PS    & 1.000  & 1.8   & 0.996  & 7.3   & 1.000  & 8.0   & 1.000  & 1.8   & 1.000  & 7.6   & 1.000  & 8.1  \\
    PLS   & 0.990  & 1.9   & 0.982  & 14.4  & 1.000  & 9.9   & 0.970  & 1.8   & 0.996  & 15.1  & 1.000  & 8.5  \\
    \midrule
    $k_2$     & CR    & AL    & CR    & AL    & CR    & AL    & CR    & AL    & CR    & AL    & CR    & AL \\
    \midrule
    IID   & 0.950  & 14.0  & 0.938  & 13.3  & 0.944  & 13.8  & 0.946  & 21.3  & 0.946  & 24.6  & 0.954  & 21.6  \\
    AR    & 0.942  & 32.3  & 0.936  & 29.0  & 0.818  & 14.7  & 0.976  & 37.7  & 0.966  & 39.4  & 0.846  & 17.6  \\
    LS    & 0.986  & 8.1   & 0.988  & 30.9  & 0.996  & 14.3  & 0.998  & 7.1   & 0.992  & 32.4  & 0.998  & 13.4  \\
    PS    & 1.000  & 1.7   & 0.996  & 11.1  & 1.000  & 12.2  & 1.000  & 1.5   & 1.000  & 11.8  & 1.000  & 12.4  \\
    PLS   & 0.746  & 6.6   & 0.962  & 34.6  & 0.946  & 22.9  & 0.734  & 7.6   & 0.982  & 40.0  & 0.914  & 24.1  \\
    \bottomrule
    \end{tabular}%
    }
  \label{tab:CI2}%
\end{table}%

\section{Real Data Analysis}\label{sec:realdata}

A dataset of inflation data from the United States spanning 1983-01 to 2023-09 is obtained from the St. Louis Fed FRED website \citep{fredcitation}. It contains 489 observed monthly percentage changes in the Consumer Price Index for All Urban Consumers: All Items in U.S. City Average. Economic data often exhibit nonstationary behavior, and the economics literature proposes regime switching models to account for shifts in inflation dynamics \citep{garcia1996analysis,https://doi.org/10.1002/1099-131X(200101)20:1<21::AID-FOR763>3.0.CO;2-0,AMISANO2013118,huber2018markov}. Changes in the mean level of inflation that occur in an economy are often explainable by economic or political factors \citep{binder2022expected}.

\begin{figure}[h!]
    \centering
    \includegraphics[width=0.9\linewidth]{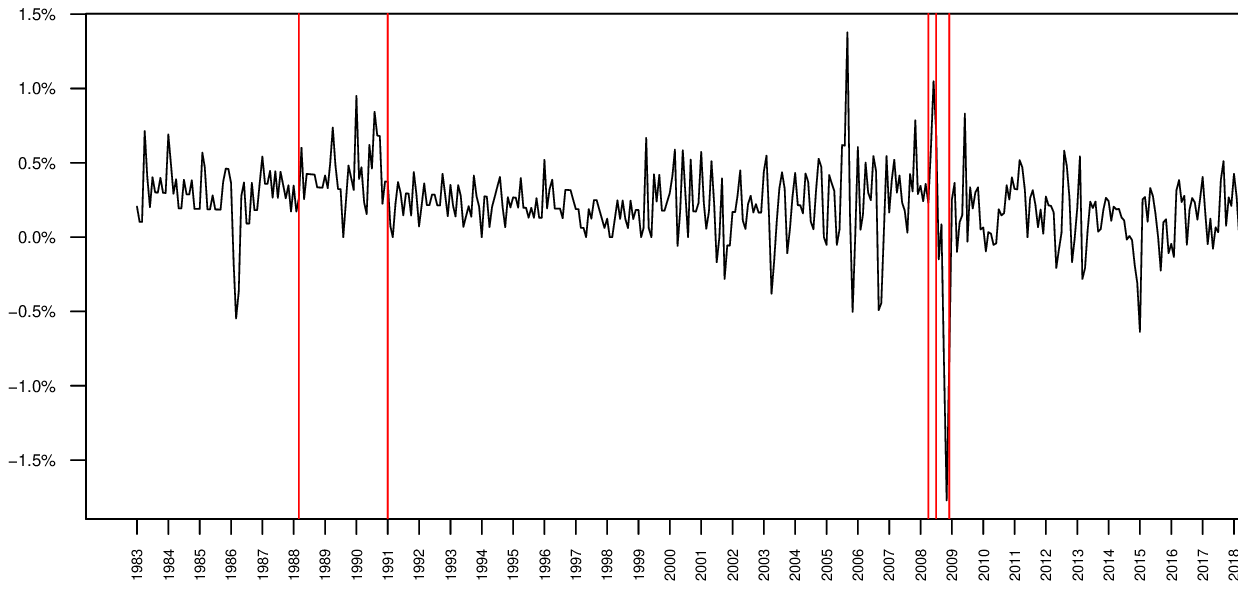}
    \includegraphics[width=0.9\linewidth]{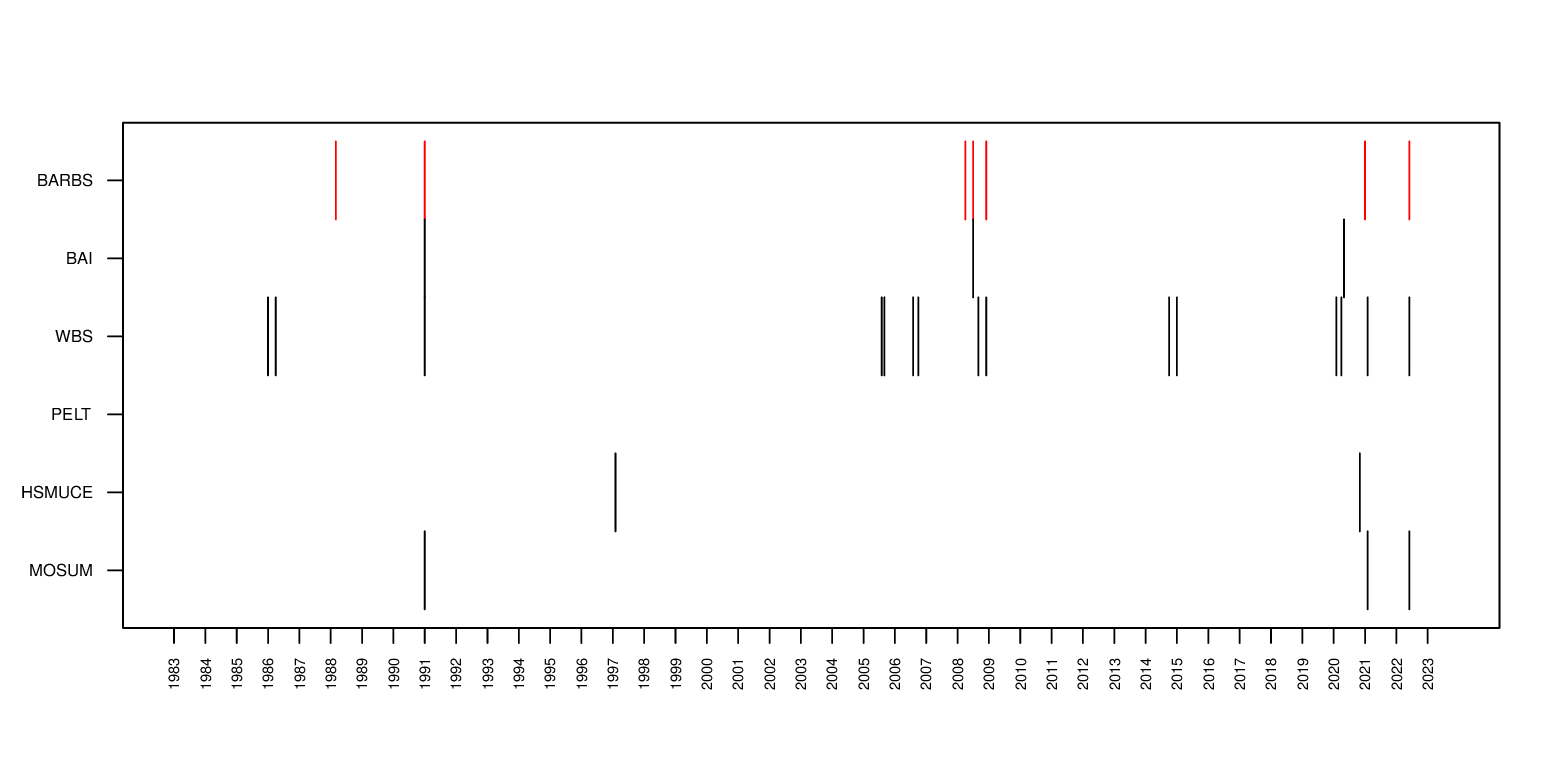}
    \caption{Upper panel:   the inflation data with change points detected by BARBS;  lower panel: comparison of the change point locations estimated by BARBS and competing methods.}
    \label{fig:inflation}
\end{figure}

As shown in Figure \ref{fig:inflation}, BARBS identifies seven change points, occurring in April 1988, February 1991, April, July, and December 2008, February 2021, and July 2022. The change around 1988-04 is detected only by BARBS among the relatively conservative procedures, whereas the change around  1991-02 is also identified by WBS and MOSUM. During 2008--2009, BARBS detects three closely spaced change points in  2008-04, 2008-07, and 2008-12; two of these locations approximately coincide with those detected by WBS, while BAI, PELT, HSMUCE, and MOSUM largely miss this episode. BARBS further identifies two changes in 2021-02 and 2022-07, both of which are also detected by WBS and MOSUM, whereas BAI and HSMUCE identify only the earlier change and miss the later one.
Overall, in this application, BARBS appears less conservative than BAI, PELT, HSMUCE, and MOSUM, which  tend to miss the changes in 1988 and the 2008–2009 period,  while avoiding the much larger number of potentially spurious detections produced by WBS.

One can use the historical perspective provided by \cite{binder2022expected} to help explain the change points detected by BARBS. 
From 1988 to 1991, U.S. inflation remained relatively high due to strong economic growth, rising wages from a tight labor market, increasing service costs, and higher energy prices caused by the Gulf crisis. In addition, the Federal Reserve’s delayed response after earlier monetary easing  in response to the 1987 stock market crash contributed to sustained inflation.

In early 1991, the end of the Gulf War eased market concerns over disruptions to global oil supply, leading to a decline in energy prices. This may have contributed to lower inflationary pressure through transportation and production costs. At the same time, the lagged effects of the Federal Reserve’s earlier tightening measures began to take hold, lowering inflation expectations and curbing spending. Besides, global economic growth slowed in the early 1990s, reducing external inflationary pressure. Events such as the collapse of Eastern European regimes and the dissolution of the Soviet Union further weakened external demand and pricing pressure.

The three changes detected in 2008 capture a brief inflation surge followed by sharp disinflation during the Great Recession.
The inflation spike in 2021–2022 is attributable to supply disruptions caused by the COVID-19 pandemic. The alignment of these major economic episodes with the detected change points illustrates the practical interpretability of BARBS within complex economic time series.

\section{Concluding remark}

This paper develops BARBS, a bootstrap-assisted robust binary segmentation framework for
multiple change point detection in nonstationary time series. Unlike traditional approaches that rely on thresholding or i.i.d. assumptions, BARBS achieves precise calibration of critical values in the presence of complex temporal dynamics, enabling asymptotic control over the probability of incorrectly identifying the number of change points at any prespecified significance level. Furthermore,  we   introduce a second-stage  procedure that yields refined estimators, and derive their asymptotic distributions and uniform convergence rate under both fixed and shrinking change magnitudes.

Possible directions for future research include incorporating BARBS into more advanced and popular variants of binary segmentation, such as  seeded binary segmentation. Another interesting extension is to adapt the BARBS framework to high-dimensional time series. In addition, the type of abrupt change points  considered in this paper may be generalized beyond mean changes to broader types of structural changes, such as distributional changes.

\bibliographystyle{apalike}
		\bibliography{ref}	
\end{document}